\documentclass[aps,twocolumn,prd,superscriptaddress,nofootinbib,showpacs,letterpaper,eqsecnum]{revtex4}

\usepackage{amsmath,amssymb}
\usepackage[dvips]{graphicx}
\usepackage{longtable}
\usepackage[usenames,dvipsnames]{xcolor}
\usepackage[normalem]{ulem}
\usepackage{bm}
\usepackage{comment}
\usepackage{xcolor}
\usepackage{soul}
\usepackage{hyperref}
\usepackage[percent]{overpic}
\usepackage{overpic}

\begin{document}

\newcommand{\bea}{\begin{eqnarray}}
\newcommand{\eea}{\end{eqnarray}}
\newcommand{\E}{\mathrm{E}}
\newcommand{\Var}{\mathrm{Var}}
\newcommand{\bra}[1]{\langle #1|}
\newcommand{\ket}[1]{|#1\rangle}
\newcommand{\braket}[2]{\langle #1|#2 \rangle}
\newcommand{\mean}[2]{\langle #1 #2 \rangle}
\newcommand{\be}{\begin{equation}}
\newcommand{\ee}{\end{equation}}	
\newcommand{\ba}{\begin{eqnarray}}
\newcommand{\ea}{\end{eqnarray}}
\newcommand{\SD}[1]{{\color{magenta}#1}}
\newcommand{\rem}[1]{{\sout{#1}}}
\newcommand{\alert}[1]{\textbf{\color{red} \uwave{#1}}}
\newcommand{\Y}[1]{\textcolor{blue}{#1}}
\newcommand{\R}[1]{\textcolor{red}{#1}}
\newcommand{\B}[1]{\textcolor{black}{#1}}
\newcommand{\C}[1]{\textcolor{cyan}{#1}}
\newcommand{\db}{\color{darkblue}}
\newcommand{\huan}[1]{\textcolor{cyan}{#1}}
\newcommand{\fan}[1]{\textcolor{blue}{#1}}
\newcommand{\ac}[1]{\textcolor{cyan}{\sout{#1}}}
\newcommand{\intinfty}{\int_{-\infty}^{\infty}\!}
\newcommand{\Tr}{\mathop{\rm Tr}\nolimits}
\newcommand{\const}{\mathop{\rm const}\nolimits}
\newcommand{\Caltech}{\affiliation{Theoretical Astrophysics 350-17, California Institute of Technology, Pasadena, CA 91125, USA}}

\title{ Stability of Force-Free Magnetospheres}

\author{Huan Yang}
\affiliation{Perimeter Institute for Theoretical Physics, Waterloo, Ontario N2L2Y5, Canada}
\affiliation{Institute for Quantum Computing, University of Waterloo, Waterloo, Ontario N2L3G1, Canada}
\author{Fan Zhang}
\affiliation{\mbox{Department of Physics, West Virginia University, PO Box 6315, Morgantown, WV 26506, USA}}

\begin{abstract}
We analyze the dynamical evolution of a perturbed force-free magnetosphere of a rotating black hole, which is described by the Blandford-Znajek solution in the stationary limit. We find that the electromagnetic field perturbations can be classified into two categories: ``trapped modes" and ``traveling waves". The trapped modes are analogous to the vacuum (without plasma) electromagnetic quasinormal modes in rotating black hole spacetimes, but with different eigenfrequencies and wave functions, due to their coupling with the background electromagnetic field and current. The traveling waves propagate freely to infinity or the black hole horizon along specific null directions, and they are closely related to the no-scattering Poynting flux solutions discovered by Brennan, Gralla and Jacobson. Our results suggest that the Blandford-Znajek solution is mode stable, and more importantly we expect this study to illuminate the dynamical behavior of force-free magnetospheres as well as to shed light on the path to new exact solutions.
\end{abstract}

\pacs{ 04.70.Bw, 94.30.cq, 46.15.Ff}

\maketitle

\section{Introduction}

It is generally believed that a magnetized, rotating black hole could power an outgoing Poynting flux (``jet") through the ``Blandford-Znajek" (BZ) process \cite{1977MNRAS.179..433B}. This process is analogous to the power-extraction process of neutron stars as discovered by Goldreich and Julian \cite{Goldreich:1969sb}, but with the outgoing flux powered by the rotational energy of the black holes instead of the neutron stars. Lately Lyutikov also discussed another power-extracting scenario with a Schwarzschild black hole moving in a constant magnetic field \cite{Lyutikov:2011}. Within the magnetosphere of the black holes, the energy density of the electromagnetic field is orders of magnitude larger than the energy density of the plasma itself, and this physical condition justifies the approximation that the plasma particles' inertia is negligible, which is often referred as the {\it force-free} approximation. It turns out that under such an approximation, 
the evolution of the electromagnetic field is self-contained, in the sense that we have a closed set of evolution equations without explicitly invoking the equations of motion for the plasma. The presence of the plasma particles instead manifests as a non-linear modification to the vacuum Maxwell equations. It is within this force-free context that the BZ solution was discovered.

Ever since the seminal work by Blandford and Znajek, much progress has been made on the study of the force-free magnetosphere of black holes.  MacDonald and Thorne \cite{MacDonald:1982zz} have formulated the force-free evolution equations in a $3+1$ language, which serves as the foundation for most of the numerical investigations later on (e.g.
 \cite{Contopoulos:1999ga,Komissarov:2002my,Asano:2005di,Cho:2004nn,
Timokhin:2006ur,Palenzuela:2010nf,Petri:2012cs,Parfrey:2011ta,
 Spitkovsky:2006np,Kalapotharakos:2008zc,Palenzuela:2010xn, 
McKinney:2006sc,Yu:2010bp,Uzdensky:2004qu,
Alic:2012df,Kalapotharakos:2011db}). Carter \cite{Carter1979} and Uchida \cite{1997MNRAS.286..931U,1997MNRAS.291..125U,1997PhRvE..56.2181U,1997PhRvE..56.2198U,1998MNRAS.297..315U} instead constructed a unified spacetime description, recasting the dynamical variables into a pair of scalar Euler potentials. 
More recently, Gralla and Jacobson \cite{Gralla:2014yja} promoted the utilization of differential forms in conjunction with the scalar potentials, and they demonstrated that this geometric language is extremely efficient at explaining previous results on exact solutions such as the family of null ($ {\bf E} \cdot {\bf B} =0$ and ${\bf E}^2= {\bf B}^2$) solutions discovered by Brennan, Gralla and Jacobson \cite{Brennan:2013jla, Brennan:2013ppa}, as well as acquiring new insights into the underlying physics. As a result, we shall also adopt the language of differential forms in our analysis.

As compared to the progress made on exact solutions, the literature on 
the dynamical perturbations and the mode structure of the force-free magnetosphere is more scarce. 
While many numerical investigations have indicated that the Blandford-Znajek process is remarkably stable against perturbations \cite{Palenzuela:2011es,Palenzuela:2010xn,
Lehner:2011aa,Komissarov:2007rc,Spitkovsky:2006np}  
a clear and precise analytical understanding of the problem is still lacking. An important step in this direction was achieved by Uchida 
\cite{1997MNRAS.286..931U,1997MNRAS.291..125U,1998MNRAS.297..315U}, who constructed a general framework for the linear perturbations of force-free systems. However, mode structure of black hole magnetospheres has never been analyzed before. In this work, we utilize the differential forms formalism to lay down an alternative set of perturbation equations. Using more recent perturbative techniques, we have been able to decouple these equations, solve them, and reveal, for the first time in literature, a remarkably clean mode structure and perturbation propagation behavior. 

\begin{figure}[t,b]\centering
\includegraphics[width=0.9\columnwidth]{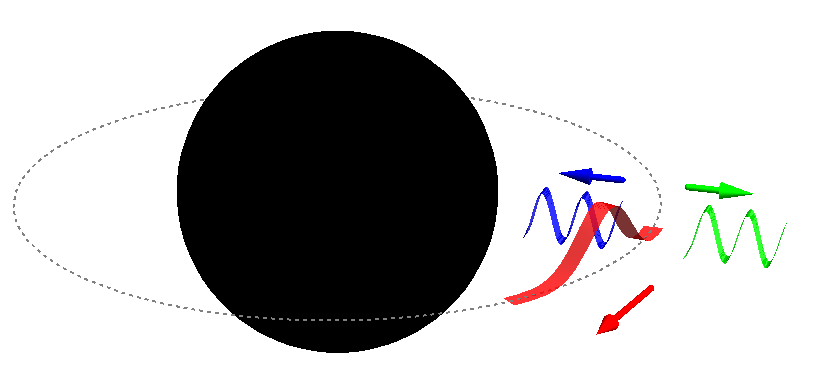}
\caption{A schematic depiction of the two classes of perturbations. The green and blue waves propagating away from and towards the black hole belong to the traveling wave class of solutions, while the red wave packet represents the trapped modes that are imperfectly confined by an effective radial potential. The black sphere in this figure represents the black hole, and the grey dashed circle represents a spherical photon orbit (for the relationship between such orbits and the trapped perturbations such as quasinormal modes, see e.g. \cite{FerrariMashhoon1984,Yang:2012he}). } 
\label{fig:TheTwoClasses}
\end{figure}

We will show that the perturbation of force-free magnetospheres fall into two classes of solutions: the ``trapped modes" and the ``traveling waves" (depicted schematically in Fig.~\ref{fig:TheTwoClasses}). The trapped modes represent a collection of electromagnetic waves (distinguished by their spherical harmonic decomposition) trapped by an effective potential, which is due to the spacetime geometry as well as the presence of the background electromagnetic field and current.
Similar to the usual black hole quasinormal modes \cite{ReggeWheeler1957,Zerilli1970b,
Teukolsky,Teukolsky:1972my,Berti2009,Yang:2013shb}, this trapping is imperfect and the modes will eventually be absorbed by the black hole or leak out to infinity. As a result, the eigenfrequencies of these modes are complex numbers (whose values we compute explicitly). On the other hand, the traveling waves comprise a class of waves that propagates along the principal null directions of the background spacetime without any backscattering. In this sense, they are closely related to the null solutions of Brennan, Gralla and Jacobson \cite{Brennan:2013jla}. An important difference here is that the solutions we find are generic perturbations to the Blandford-Znajek solution, while the whole spacetime is still magnetically dominated
\footnote{By magnetic dominance, we mean that ${\bf B}^2 > {\bf E}^2$ everywhere in the plasma.},
and the current density is generically not null. It is also important to note that this non-scattering mode family is a new feather of perturbations of force-free plasma, which is not present in vacuum. While analyzing these two classes of solutions, we observe no unstable modes, thus providing an affirmative answer to mode stability, and initiating the first step towards a proof of full non-linear stability. We also note that the understanding gained here of the perturbing modes, such as their characteristic frequencies, will help interpret astronomical observations of the ``settling-down'' stage of perturbing events, enabling in particular the extraction of important information regarding the host black hole. 


We note that because the Blandford-Znajek solution is only known in the small black-hole spin limit, we shall only focus on slowly spinning black holes. In addition, the perturbative analysis here is performed on the monopole solution, which describes the local perturbations (on the northern or southern hemisphere) of the split-monopole solution. In order to obtain a global solution, the solution on the northern and southern hemispheres have to be matched on the current sheet, which is also the equatorial plane here.
The structure of this paper is organized as follows: We begin by introducing the general framework for generating perturbation equations in Sec.~\ref{sec2}, before specializing to the perturbations of the magnetic monopole solution in a Schwarzschild spacetime in Sec.~\ref{sec3}. Equipped with the intuition and results obtained in Sec.~\ref{sec3}, we tackle the perturbations to the Blandford-Znajek solution in Sec.~\ref{sec4}. Final concluding remarks are presented in Sec.~\ref{sec5}.

\section{Formalism}\label{sec2}
In this section, we set up the basic framework for our perturbative study. As mentioned in the introduction, we shall rely heavily on differential forms in the following analysis, as promoted in Ref.~\cite{Gralla:2014yja}. A useful review of differential forms can be found in Appendix A of \cite{Gralla:2014yja}, or textbooks on differential geometry.

For later convenience, we quote the Kerr metric in Boyer-Linquist coordinates for a black hole with mass $M$ and spin $a$, and we are particularly interested in its small-spin limit (to first order in $a$):
\begin{equation} \label{eq:slowspinmetric}
d s^2= -\frac{\Sigma \Delta}{A} dt^2+\frac{A}{\Sigma}\sin^2\theta (d \phi-\Omega_Z dt)^2+\frac{\Sigma}{\Delta} dr^2+\Sigma d\theta^2\,,
\end{equation}
whereby the quantities appearing in the metric (and their small-spin limits) are 
\begin{align}
\Sigma = & r^2+ a^2\cos^2 \theta \approx r^2,\,\,\, \frac{\Delta}{r^2} = 1-\frac{2M}{r}+\frac{a^2}{r^2} \approx \left (1-\frac{2M}{r} \right )\nonumber \\
A = & (r^2+a^2)^2-a^2\Delta \sin^2\theta \approx r^4, \,\,\, \Omega_Z =\frac{2 M a r}{A} \approx \frac{2 M a}{r^3}\,.
\end{align}
The black hole's outer horizon is located at $r_+ = M +\sqrt{M^2-a^2}$ which reduces to $2M$ in the Schwarzschild limit. In addition, another commonly used radial (``tortoise") coordinate $r_*$ is defined through $dr_* = (r^2+a^2)/\Delta dr$. For the rest of the paper, we will assume $M=1$ whenever explicit numerical calculations are performed, such as when generating the figures. 

In the force-free plasma, the electromagnetic field is degenerate:$F \wedge F=0$, where $F$ is the field $2$-form. Together with the source-free part of the Maxwell equation $d F =0$, it can be shown \cite{Carter1979,1997MNRAS.286..931U,1997MNRAS.291..125U,
1997PhRvE..56.2181U,1997PhRvE..56.2198U,1998MNRAS.297..315U} 
that there is always at least one pair of ``Euler potentials" $\phi_1, \,\phi_2$ such that
\begin{equation}
F = d \phi_1 \wedge d \phi_2\,.
\end{equation}  
The force-free condition further requires that 
\begin{equation}\label{eqffecon}
d \phi_1 \wedge d * F=0,\quad d \phi_2 \wedge d * F=0\,,
\end{equation}
where $*$ stands for the Hodge dual operator. Note ``Euler potentials" are \emph{locally} defined functions that depend on the spacetime coordinates, and they are not unique for a given field $2$-form $F$. This gauge freedom can be fixed by specifying the expressions for $\phi_1,\, \phi_2$, and imposing appropriate boundary conditions. In particular, the Blandford-Znajek solution can be represented by
\begin{equation}\label{eqbzsol}
\phi_1 = q \cos\theta,\quad \phi_2= \phi - \frac{\Omega_H}{2} u\,.
\end{equation}
Here $q$ is the monopole charge which can be determined by integrating the total magnetic flux over the northern/southern hemisphere, $\Omega_H=a/(r^2_++a^2) \approx a/(4 M^2)$ and $u=t-r_*$ is the outgoing tortoise coordinate. Using Eq.~(\ref{eqbzsol}), it is straightforward to check that the above solution satisfies the force-free condition in Eq.~(\ref{eqffecon}) up to $\mathcal{O}(a)$ order.

Now suppose that the force-free magnetosphere as described by the Blandford-Znajek solution is perturbed from its original state. The new field $2$-form is expressed by a new pair of ``Euler potentials" $\tilde \phi_{1,2}$:
\begin{equation}\label{eqfieldexpan}
\tilde F = F +\epsilon \delta F = d \tilde \phi_1 \wedge d \tilde \phi_2\,,
\end{equation}
where the perturbative expansion parameter $\epsilon \ll 1$. As the perturbation magnitude is small, it is always possible to find the pair of $\tilde \phi_1$ and $\tilde \phi_2$ such that
\begin{equation}\label{eqphitilde}
\tilde \phi_1 = \phi_1 + \epsilon \alpha,\quad \tilde \phi_2 = \phi_2+\epsilon \beta\,,
\end{equation}
where $\alpha, \beta$ are the perturbative ``Euler potentials". The perturbative field $2$-form can then be read off from Eqs.~(\ref{eqfieldexpan}) and (\ref{eqphitilde}) as 
\begin{equation} \label{eq:PertFaraday}
\delta F = d \alpha \wedge d \phi_2+ d \phi_1 \wedge d \beta\,,
\end{equation}
and the force-free condition for the total field $\tilde{F}$ can be translated into (up to $\mathcal{O}(\epsilon)$ order)
\begin{align}\label{eqgffec}
&d \alpha \wedge d * F+ d \phi_1 \wedge d * \delta F = 0, \nonumber \\
&d \phi_2 \wedge d * \delta F+d \beta \wedge d * F=  0\,.
\end{align}
We can see that the above expressions contain both the coupling between the perturbative current $d * \delta F$ and the background field as represented by $d\phi_1$ and $d\phi_2$, and the coupling between the perturbative fields and the background current $d * F$. It can be applied to generic backgrounds as long as the background ``Euler potentials" are known. The perturbative electromagnetic field with these couplings has a richer structure than the perturbations in a vacuum Kerr spacetime, as will be explained in more details in later sections. 

\section{Perturbations of the Schwarzschild monopole solution}\label{sec3}

In order to gain some physical intuition before attacking the full problem, we first study the perturbation of the Schwarzschild monopole solution. For simplicity, we will set the monopole charge to $1$, in which case the field $2$-form is just
\begin{equation}
F =d \cos \theta \wedge d \phi\,.
\end{equation}
There is no background current in this solution ($d * F=0$), and the force-free conditions reduce to  
\begin{equation} \label{eq:SFFCond}
  d \cos \theta \wedge d * \delta F =0, \quad d \phi \wedge d * \delta F=0\,,
 \end{equation}
 with the perturbed field $2$-form given by
 \begin{equation}\label{eqSdeltaF}
 \delta F = d \cos \theta \wedge d \beta + d \alpha \wedge d \phi\,.
 \end{equation} 
 Combining Eqs.~(\ref{eq:SFFCond}) and (\ref{eqSdeltaF}), it is then straightforward to derive the following wave equations for $\alpha$ and $\beta$ (defining $x \equiv \cos \theta$):
 \begin{align}
&\left (-\frac{\partial^2}{\partial t^2} +\frac{\partial^2}{\partial r^2_*}\right ) \alpha \notag \\
& \quad\quad +(1-x^2) \frac{1-2M/r}{r^2}\left( \frac{ \partial^2 \alpha}{ \partial x^2}+\frac{\partial^2 \beta}{\partial x \partial \phi} \right) =0 \label{eq:SchCouple1} \\
&\left (-\frac{\partial^2}{\partial t^2}  +\frac{\partial^2}{\partial r^2_*}\right ) \beta+\frac{1-2M/r}{r^2(1-x^2)}\left( \frac{\partial^2 \beta}{\partial \phi^2}+\frac{\partial^2 \alpha}{\partial \phi \partial x}\right)=0\, \label{eq:SchCouple2}.
\end{align}
Even when assuming a separable form for the solutions with $\phi$ dependence of $\alpha$ and $\beta$ given by $e^{im\phi}$, these two equations are generally coupled when $m \neq 0$. In the rest of this section, we will decouple the equations via a ``basis transformation'' and discover that there are two distinctive families of solutions (the ``tapped modes" and the ``traveling waves"). We discuss their behaviours separately in Sections \ref{sec:SchTrap} and \ref{sec:SchTrav}, respectively.
 
\subsection{Trapped modes \label{sec:SchTrap}}
Let us now define a new basis variable
\begin{equation}\label{eqt1}
\gamma_1 \equiv \partial_x \alpha + \partial_{\phi} \beta \,.
\end{equation} 
Then taking $\partial_x$ against Eq.~\eqref{eq:SchCouple1} and $\partial_{\phi}$ against \eqref{eq:SchCouple2} before adding up the two expressions leads to a decoupled wave equation for trapped modes \footnote{As explained below, the radial wave equation contains an potential term which effectively traps the mode excitations, so that we refer these modes as ``trapped modes".}:
 \begin{align} \label{eq:DiaSch1}
 &\left (-\frac{\partial^2}{\partial t^2}  +\frac{\partial^2}{\partial r^2_*}\right ) \gamma_1+\frac{1-2M/r}{r^2}\left \{ \frac{\partial}{\partial x} \left [(1-x^2) \frac{\partial \gamma_1}{\partial x}\right ]\right \} \nonumber \\
 &\quad\quad+\frac{1-2M/r}{r^2(1-x^2)} \frac{\partial^2 \gamma_1}{\partial \phi^2}=0\,.
 \end{align}
One can easily check that Eq.~(\ref{eq:DiaSch1}) is separable under the ansatz $\gamma_1= e^{-i \omega t} e^{i m \phi}\chi_1(r ) P^m_l(x)$, where $P^m_l(x)$ is the associated Legendre polynomial. In addition, the radial equation for $\chi_1(r)$ can be deduced from Eq.~\eqref{eq:DiaSch1} as
\begin{equation} \label{eq:Chi1}
\frac{d^2}{d r_*^2}\chi_1(r) + \left[ \omega^2 -\frac{l(l+1)(1-2M/r)}{r^2}\right]\chi_1(r) = 0\,,
\end{equation}
which is the standard Regge-Wheeler-type wave equation in Schwarzschild spacetime \cite{ReggeWheeler1957,Zerilli1970b,PhysRev.97.511,Detweiler1976} for electromagnetic perturbations. After imposing the boundary conditions that the wave is ingoing at horizon and outgoing at spatial infinity, 
\begin{equation}
\label{eqSasym}
\chi_1(\omega,r) \equiv \left\{
\begin{array}{c}
e^{-i\omega r_*}\,, 
 \qquad \qquad r_*\rightarrow-\infty\, \\
\\
e^{i\omega r_*}\,, \qquad \qquad  r_* \rightarrow \infty\,,
\end{array}\right.
\end{equation}
we can solve the eigenvalue problem and obtain the discrete set of $\omega_{l}$ values as the quasinormal mode frequencies. The relationship between this equation and the Teukolsky equation with which the quasinormal modes of Newman-Penrose quantities \cite{Newman1962} are usually computed is given in Ref.~\cite{Detweiler1976}. Once we have solved for $\gamma_1$, we can use the expressions in Appendix \ref{sec:AlphaBetaFromGamma} to recover $\alpha$ and $\beta$, and subsequently $\delta F$.

The equivalence between the trapped modes and the vacuum electromagnetic quasinormal modes suggests that in the Schwarzschild spacetime, the trapped mode perturbations also generate no current. In this sense, the force-free conditions in Eq.~(\ref{eq:SFFCond}) are trivially satisfied, and the wave equations can instead be derived from the vacuum Maxwell equations:
\begin{equation}
 d\, \delta F=0\,, \quad d* \delta F=0\,.
\end{equation}
This is possible because the background monopole solution has no current, so we are at liberty to add any current-free (thus vacuum, but note that not all vacuum solutions satisfy the force-free equations and constraint $\bf{E}\cdot \bf{B}=0$) perturbative solution to it, without needing to be concerned with field-current interactions. This property of the trapped modes in the Schwarzschild spacetime is however no longer true if the black hole is rotating, in which case we shall find a non-vanishing current generated by the trapped mode perturbations, and the current-field coupling is important there.

\subsection{traveling waves \label{sec:SchTrav}}

The other family of solutions can be obtained by defining
\begin{equation}\label{eqt2}
\gamma_2 \equiv (1-x^2)\partial_x [(1-x^2)\beta]-\partial_\phi \alpha
\,,
\end{equation}
and take $\partial_{\phi}$ against Eq.~\eqref{eq:SchCouple1} and $\partial_{x}$ against Eq.~\eqref{eq:SchCouple2} multiplied by $1-x^2$. After taking the difference between the results (with another $1-x^2$ multiplied onto Eq.~\eqref{eq:SchCouple2}), we obtain
\begin{equation} \label{eq:DiaSch2}
 \left ( -\frac{\partial^2}{\partial t^2} +\frac{\partial^2}{\partial r^2_*}\right ) \gamma_2=0\,.
\end{equation} 
Obviously the solutions in the time domain are $\gamma_-(t-r_*,\theta,\phi)$ and $\gamma_+(t+r_*,\theta,\phi)$, which describe wave packets propagating freely along the ingoing or outgoing principal null direction, and hence we refer to them as the ``traveling waves". Unlike the quasinormal modes or the trapped modes, these traveling waves don't experience any scattering from a potential, which is a property shared by the null solutions discovered by Brennan, Gralla and Jacobson \cite{Brennan:2013jla}. In fact, it is easy to check that the perturbative solution $\delta F$ here is the Schwarzschild limit of their null solutions, although here we have written them in a more explicit form. As was noted in Ref.~\cite{Brennan:2013jla}, the background monopole, in addition to being current-less, also does not interact with the perturbative current of the null solutions,   
so that the null solutions adding to the Schwarzschild monopole solution still satisfies the force-free condition. It is also important to note that these observations are no longer true for traveling waves in force-free plasma surrounding a rotating black hole, where we see couplings between the perturbative field and the background current, as well as forces by the background field acting on the perturbative current.

\section{Perturbations of the Blandford-Znajek solution \label{sec4}}

If the back hole is rotating, Blandford and Znajek showed that the black hole drags the magnetic field lines to co-rotate at half of the horizon frequency. During this process, the black hole gradually loses its rotational energy, which is carried away as outgoing Poynting flux to infinity. As mentioned in Sec.~\ref{sec2}, the electromagnetic field here can be described by the Blandford-Znajek solution in Eq.~(\ref{eqbzsol}), where solutions with positive and negative monopole charge $q$ should be attached at the equatorial plane.

The perturbations of the Blandford-Znajek solution have to satisfy the force-free conditions described by Eq.~(\ref{eqgffec}). Similar to the Schwarzschild monopole background case, they give rise to two coupled wave equations in the following form:
\begin{align}
(\mathcal{Q}_1 \alpha &+\mathcal{H}_1 \beta)\, \mathcal{E} \equiv \nonumber \\
&d \alpha \wedge d * F+ d \cos \theta \wedge d * \delta F= 0\,,  \label{eqwaveab1}\\
(\mathcal{H}_2 \alpha&+\mathcal{Q}_2 \beta)\, \mathcal{E} \equiv \nonumber \\
&(d \phi -\Omega_H/2 du)\wedge d * \delta F+d \beta \wedge d * F = 0\,,\label{eqwaveab2} 
\end{align}
where $\mathcal{Q}_i$ and $\mathcal{H}_i$ $i\in \{1,2\}$ are differential operators, and $\mathcal{E} = d\theta\wedge d\phi \wedge dr \wedge dt$. The detailed expressions for $\mathcal{Q}_i$ and $\mathcal{H}_i$ are listed in Appendix \ref{sec:QandH}. Note that similar to Blandford and Zanjek \cite {1977MNRAS.179..433B}, we only keep terms up to $\mathcal{O}(a)$ order, and the force-free condition in Eq.~(\ref{eqgffec}) is also satisfied to $\mathcal{O}(a)$ order. A more general treatment requires first knowing the exact form of Blandford-Znajek-type solutions for generic black hole spins, which is currently unavailable.

In order to solve the coupled wave equations, we shall transform them from the $(\alpha, \beta)$ basis to the $(\gamma_1, \gamma_2)$ basis, under which they are decoupled in the Schwarzschild limit. Applying Eqs.~(\ref{eqt1}) and (\ref{eqt2}) to Eqs.~(\ref{eqwaveab1}) and (\ref{eqwaveab2}), and after a lengthy but nevertheless straightforward calculation, we arrive at the wave equations in terms of the new basis variables:
\begin{align}
& \mathcal{H}_V \gamma_1+ a \mathcal{V}_1 \gamma_1+ a \mathcal{V}_2 \gamma_2=0\,,  \label{eqwg1g21} \\
& \mathcal{H}_P \gamma_2+ a \mathcal{P}_1 \gamma_1+ a \mathcal{P}_2 \gamma_2=0\,, \label{eqwg1g22}
\end{align}
where the detailed expressions for operators $ \mathcal{H}_V, \mathcal{H}_P$, $ \mathcal{V}_{1,2}$ and $\mathcal{P}_{1,2}$ are given in Appendix \ref{sec:PerEqGamma}, and they are all devoid of $a$ dependence. Subsequently, $a \mathcal{V}_{1,2}$ and $a \mathcal{P}_{1,2}$ are proportional to the black hole spin $a$, and in the slow-rotation limit, they can be treated as components of a perturbative ``Hamiltonian" in the wave equations. For Schwarzschild black holes with $a=0$, the perturbative Hamiltonian vanishes, and the above wave equations reduce to Eqs.~(\ref{eq:DiaSch1}) and (\ref{eq:DiaSch2}). In the next section, we shall introduce a perturbation technique to solve Eqs.~(\ref{eqwg1g21}) and \eqref{eqwg1g22}.

\subsection{Perturbative method for solving coupled wave equations}

As $ \mathcal{H}_V, \mathcal{H}_P$ are independent of $a$, the Schwarzschild wave equations (\ref{eq:DiaSch1})  and (\ref{eq:DiaSch2}), or
\begin{equation}
 \mathcal{H}_V \gamma^{(0)}_1=0,\quad \mathcal{H}_P \gamma^{(0)}_2=0\,,
\end{equation}
give the solutions described in Sec.~\ref{sec3}, with the eigenfrequencies $\omega_S$ being the same as vacuum quasinormal-mode frequencies. Let's examine the spin induced perturbations to the trapped modes first, and add the perturbative Hamiltonian back into the wave equations and consider Eq.~(\ref{eqwg1g21}) in full. 
We can expand the mode wave functions as
\begin{equation}\label{eqwaveexpan}
\gamma_1 = \gamma^{(0)}_1 + a \gamma^{(1)}_1+\mathcal{O}(a^2),\quad \gamma_2 = a \gamma^{(1)}_2+\mathcal{O}(a^2)\,,
\end{equation}
where we note in particular that at the zeroth order in $a$, the trapped modes have no components in the $\gamma_2$ sector. 
Plugging Eq.~(\ref{eqwaveexpan}) into the Eq.~(\ref{eqwg1g21}), and by going into the frequency domain, we find that Eq.~(\ref{eqwg1g21}) becomes
\bea\label{eqp1}
0&=&\mathcal{H}_V (\omega_S+\delta \omega) [\gamma^{(0)}_1+ a \gamma^{(1)}_1]+a \mathcal{V}_1 \gamma^{(0)}_1+\mathcal{O}(a^2)\,\nonumber\\
&\approx& a\mathcal{H}_V (\omega_S)  \gamma^{(1)}_1+ \left .\frac{\partial \mathcal{H}_V}{\partial \omega} \right |_{\omega_S} \delta \omega\,  \gamma^{(0)}_1+ a \mathcal{V}_1 \gamma^{(0)}_1\,.
\eea
Note that now this equation has naturally decoupled from $\gamma_2$ at $\mathcal{O}(a)$ order, which is the reason why we switch to the $(\gamma_{1},\gamma_{2})$ basis in this analysis.

In order to solve the above equation and obtain the frequency shift $\delta\omega$, we shall apply a technique that Mark, Yang, Zimmerman, and Chen first developed for studying perturbations to Kerr-Newman black holes \cite{KerrNewman, Zimmerman2014pr} and was later applied to near-extremal black holes to investigate turbulent-like instabilities \cite{Yang:2014tla}. In particular, these studies have verified the validity of this method under a variety of different scenarios. For example, one can successfully compute the quasinormal mode frequencies in slowly spinning Kerr black hole spacetimes with it, starting from the frequencies in the Schwarzschild spacetime. 

\begin{figure}[t,b]\centering
\includegraphics[width=0.9\columnwidth]{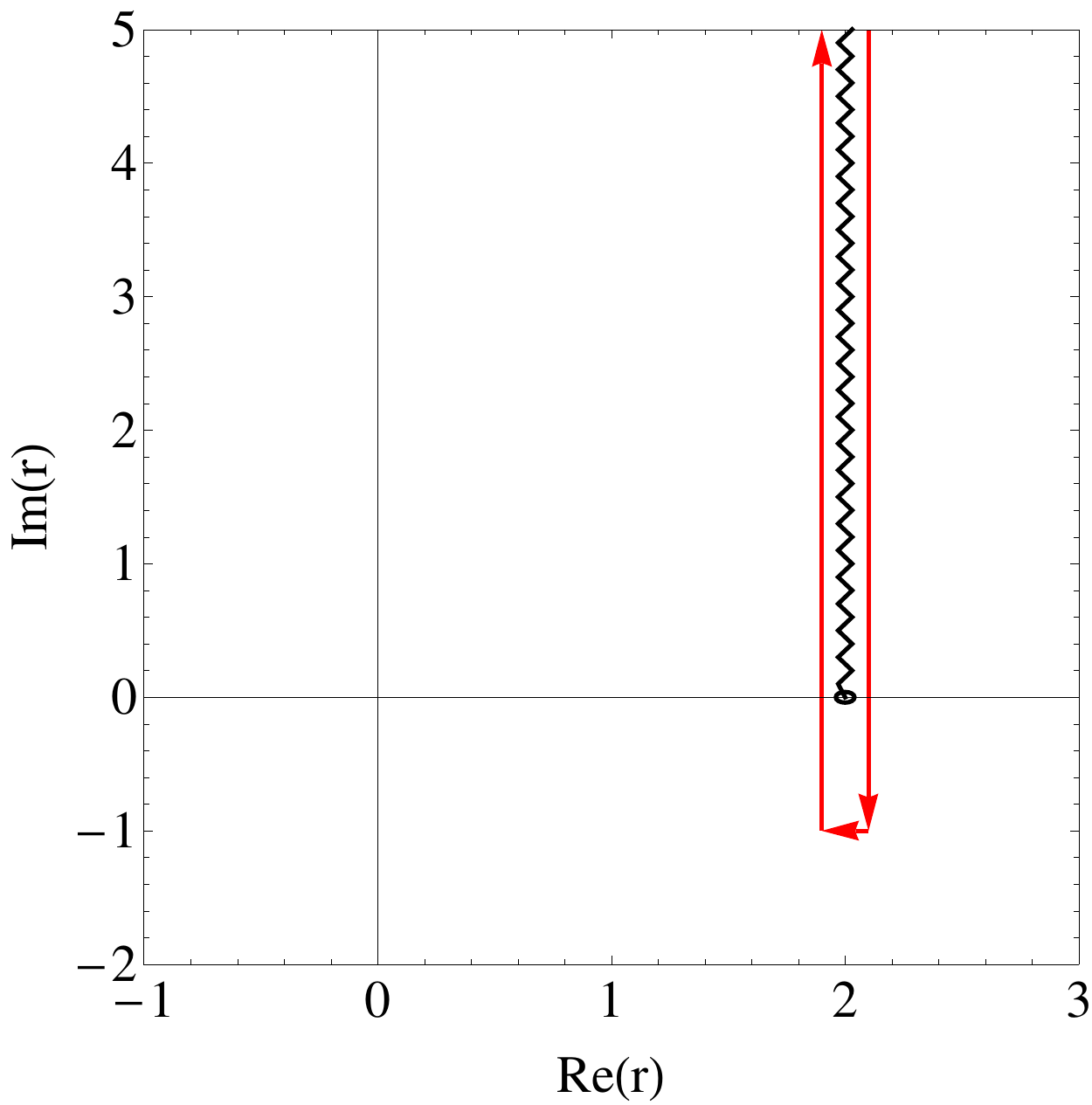}
\caption{An illustration of the contour (the red curve) in the complex $r$ plane with which we compute the inner product. The Schwarzschild wave function is analytical everywhere except at the brach cut (shown as a zigzag line), which starts at $r=2M$ and shoots upward to $+i \infty$.} 
\label{fig:contour}
\end{figure}

The central ingredient is an inner-product against which $\mathcal{H}_V$ is a self-adjoint operator
\begin{equation}
\langle \psi | \mathcal{H}_V \eta \rangle = \langle \mathcal{H}_V \psi | \eta \rangle \,,
\end{equation}
for arbitrary trapped mode wave functions $\psi$ and $\eta$\,. According to the expression of $\mathcal{H}_V$ that can be read off directly from Eq.~(\ref{eq:DiaSch1}), one such inner-product is given by
\begin{equation}
\langle \psi | \eta \rangle = \int_{\mathcal{C}} \frac{d r}{1-2M/r} \int_{-1}^1 d \cos \theta\, \psi\, \eta\,,
\end{equation}
where the integration for $r$ is along a contour $\mathcal{C}$ in the complex $r$ plane, which is graphically depicted in Fig.~\ref{fig:contour}. One can easily check that the boundary conditions in Eq.~(\ref{eqSasym}) guarantee that this contour integration produces a finite result for arbitrary trapped mode wave functions $\psi$ and $\eta$. Equipped with the inner product, we can then multiply Eq.~(\ref{eqp1}) by $\gamma^{(0)}_1$ using the ``left product", and obtain:
\begin{equation} \label{eq:FreqShift}
\delta \omega = -\frac{a\langle \gamma^{(0)}_1 | \mathcal{V}_1 \gamma^{(0)}_1 \rangle}{2\omega_S \langle \gamma^{(0)}_1 | \gamma^{(0)}_1 \rangle}\,.
\end{equation}
where we have substituted in $\partial \mathcal{H}_V/\partial \omega|_{\omega_S} = 2\omega_S$. 

For traveling waves, we can use similar arguments and conclude that the expression
\begin{equation}\label{eqtrav}
\mathcal{H}_V \gamma_2+ a \mathcal{P}_2 \gamma_2 \approx0
\end{equation}
already captures the physics up to $\mathcal{O}(a)$ order. In the following sections, we shall solve the perturbations in the trapped and traveling sectors explicitly, using the method outlined above. We shall also discuss the physical implications of these results.

\subsection{Trapped modes}

The trapped modes in the Schwarzschild limit are identical to the vacuum electromagnetic quasinormal modes, and their wave functions can be written as $\gamma^{(0)}_1= e^{-i \omega_S t} e^{i m \phi}\chi_1(r ) P^m_l(x)$. The explicit expressions for the radial wave function $\chi_1(r)$ can be found in Ref.~\cite{Leaver1985} in an expansion form, which can be inserted into Eq.~(\ref{eq:FreqShift}) to compute the frequency shifts. 

\begin{figure}[t,b]
\begin{overpic}[width=0.95\columnwidth]{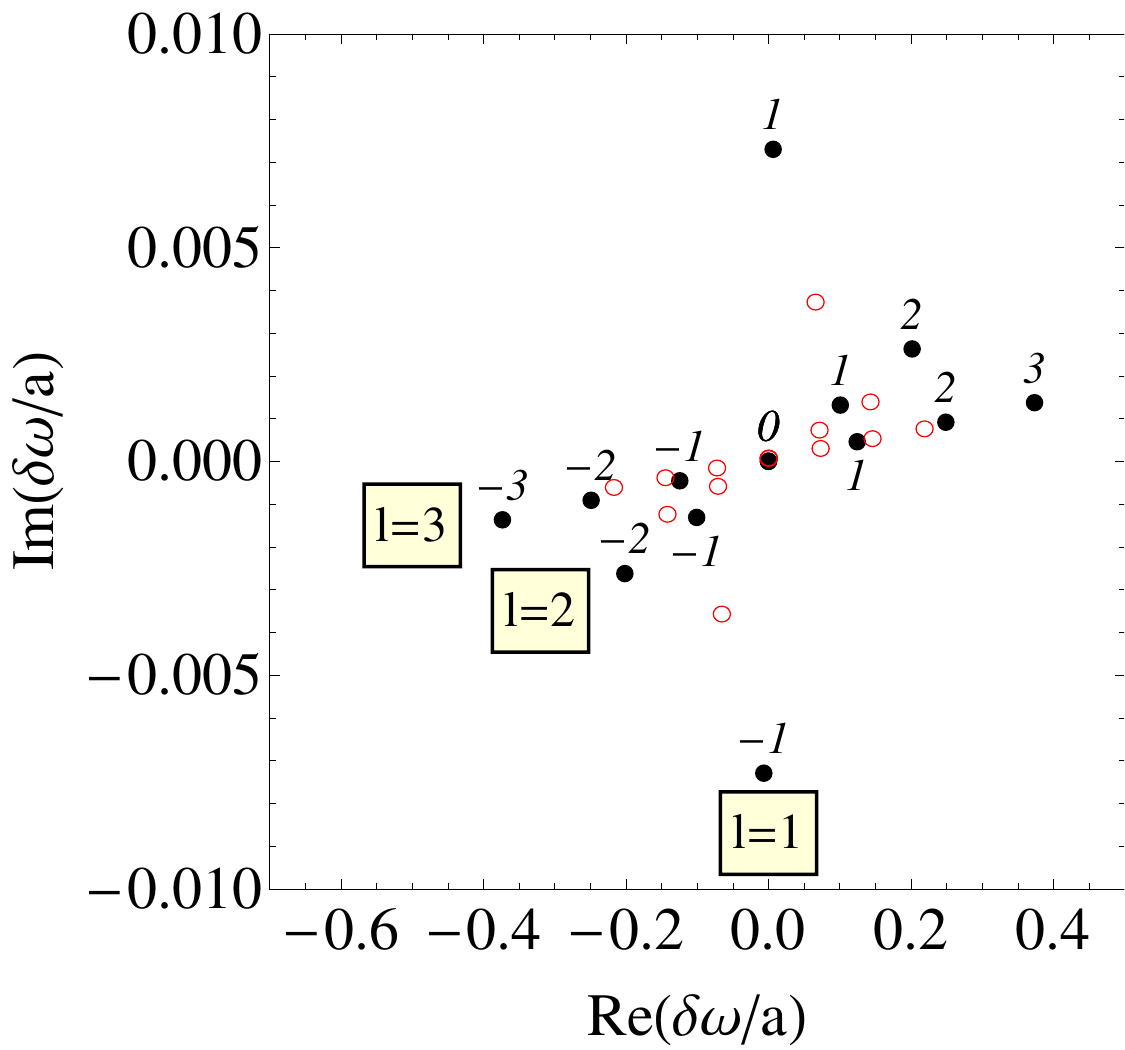}
\end{overpic}
\caption{ The black dots represent $\delta \omega/a$ for the frequency shifts of the FF trapped modes in the complex plane, where we have labelled the lines of points sharing the same $l$ (on straight lines through the origin), while the $m$ values are placed near the data points. For comparison, we also show as red circles, the corresponding $\delta \omega/a$ values for the vacuum QNMs. Their trends are the same as the FF modes and so we have not labelled them explicitly.
}
\label{fig:FreqShift}
\end{figure}

Some features of the results are immediately obvious. For example, as $\omega_S$ is independent of $m$, $\chi_1$ satisfies an equation (Eq.~\ref{eq:Chi1}) that has no $m$ dependence, and subsequently the denominator in Eq.~\eqref{eq:FreqShift} is independent of $m$. In addition, as $\mathcal{V}_1 \propto m$ (see Eq.~\ref{eq:Integrands}), we have $\delta \omega =0$ when $m=0$, and $\delta \omega$ has a linear dependence on $m$ otherwise. The results for $\delta\omega/a$ is plotted in Fig.~\ref{fig:FreqShift}. For comparison, we also show the $\delta\omega^{\text{VAC}}/a$ values for vacuum electromagnetic quasinormal modes in Fig.~\ref{fig:FreqShift}, computed as the difference between the frequencies in a Kerr spacetime with $a=0.01$ and those in a Schwarzschild spacetime. The two systems have different sets of frequency shift values, with the force-free modes demonstrating a more pronounced shift. However, the general trends in both cases are similar, and we see that the frequency shifts for both systems share the same linear dependence on $m$, and $\delta \omega =0$ when $m=0$ is also true in both cases. 

Although these results are obtained in the slow rotation limit, we can nevertheless naively linearly extrapolate the results to $a=1$ and we find no sign of mode instabilities. This observation is consistent with conclusions from previous numerical investigations 
\cite{Palenzuela:2011es,Palenzuela:2010xn,
Lehner:2011aa,Komissarov:2007rc,Spitkovsky:2006np}, which have all shown the Blandford-Znajek solution to be robust against perturbations. At this point, an interesting question to ask is that since we know near-extremal Kerr black holes possess a family of slowly-damped modes (with decaying rates $\propto \sqrt{1-a} $) 
\cite{Yang:2012he,
Yang:2012pj,Yang:2013uba}, would there also be a family of slowly-damped force-free trapped modes, or even unstable trapped modes in the $a \rightarrow 1$ limit? To answer this question, of course we need to understand the Blandford-Znajek-type solution in the near-extremal spin limit first, which will be subject to future investigation.

\subsection{traveling waves}

With the traveling waves satisfying Eq.~\eqref{eqtrav}, we can separate out the angular dependence and choose the ansatz for the wave function to be
\begin{equation}\label{eqg2c2}
\gamma_2 = e^{i m \phi} (1-x^2) P^m_l(x) \chi_2(t,r)\,,
\end{equation}
where the extra $(1-x^2)$ factor is introduced to ensure a simple relationship between $\gamma_2$ and $\alpha$ and $\beta$ (see Appendix \ref{sec:AlphaBetaFromGamma} for details). 
Combining Eq.~(\ref{eqg2c2}) and Eq.~\eqref{eq:P2}, we can derive the following wave equation:
\begin{equation}
\frac{\partial^2 \chi_2}{\partial r^2_*}-\frac{\partial^2 \chi_2}{\partial t^2}-\left [ 1-\frac{1}{l(l+1)}\right ]\frac{i m a}{4M^2} \left(\frac{\partial}{\partial t} +\frac{\partial}{\partial r_*}\right )\chi_2=0\,,
\end{equation}
The solution of this wave equation can be written as
\begin{align}
&\chi^-_2 = f_-(t-r_*)\exp\left[{i \frac{m a}{4M^2} \left(1-\frac{1}{l(l+1)}\right)r_*}\right]\,, \nonumber \\
&\chi^+_2= f_+(t+r_*)\,.
\end{align}
Aside from the extra phase factor on $\chi^-_2$, we see that these solutions still describe waves propagate along null directions without back-scattering (change of propagation direction). 

\section{Conclusion \label{sec5}}

We have analysed the perturbative behavior of a force-free magnetosphere surrounding a rotating black hole, which has a non-vanishing magnetic monopole-like charge.  We find that the perturbative electromagnetic field can generically be classified into two families: the ``trapped modes" and the ``traveling waves". The trapped modes are very similar to vacuum electromagnetic quasinormal modes of rotating black holes (in the Schwarzschild limit they are the same), but with different eigenfrequencies and wave functions. More importantly, these trapped modes possess non-vanishing charge and current if the host black hole is rotating, and are ``genuine'' force-free solutions. The other class of perturbations, the traveling waves, originates from the collective excitation of the electromagnetic field and currents. We have demonstrated the non-scattering nature of these waves in the slow-rotation limit, but we expect this property to carry over to the rapidly-spinning-black-hole scenario. 

At this point, let us emphasize the point why we have found an extra family of perturbations as compared to the vacuum case, even though we are still examining the electromagnetic field.
The key lies in the plasma-supported charge and current in the spacetime. Although the plasma equations of motion are invisible in the force-free formalism, its presence nevertheless activates the source part of the Maxwell equations, and provides some flexibility that allows for the traveling modes. 

In order to translate our results to astronomically more realistic split-monopole solutions, we have to solve the perturbative fields in the northern and southern hemispheres (with opposite monopole charges) separately, and attach the solutions at the equatorial plane. In practice, given the solution on one hemisphere, the perturbations on the opposite hemisphere can be obtained by switching $\epsilon \rightarrow-\epsilon$, with all the previous conclusion remain unchanged. Based on our results, we expect that if we initially apply some perturbation to the force-free magnetosphere, such as any electromagnetically loud astronomical event occurring in the vicinity of a supermassive black hole, and if the initial perturbing field can be decomposed into trapped and traveling modes, then at later times the perturbation would either fade away after ringing down, propagate to infinity, or be absorbed by the black hole. In the end, the magnetosphere should settle down to its original state. If a distant observer catches some of the ringdown signals emitted during this process, an inference of the spin of the black hole can be made by analyzing the frequencies present in the signal.

In the future, we expect the understanding of the perturbative behavior of black hole force-free magnetospheres developed in this work to be verified by numerical experiments. In addition, with help from numerical tools, it should be possible to extend this work to generic black hole spins.

\acknowledgements
We are grateful for many helpful comments from the anonymous referee. We thank Luis Lehner for advices on various aspects of this work, and Ted Jacobson for discussions about force-free plasma in general. We also thank Aaron Zimmerman and Zachary Mark for validating the contour integration technique in a separate study. HY acknowledges supports from the Perimeter Institute of Theoretical Physics and the Institute for Quantum Computing. Research at Perimeter Institute is supported by the government of Canada and by the Province of Ontario though Ministry of Research and Innovation.

\appendix 

\section{Recovering $\alpha$ and $\beta$ from $\gamma_1$ and $\gamma_2$ \label{sec:AlphaBetaFromGamma}}
Once we obtain the solutions for $\gamma_1$ and $\gamma_2$, we will need to recover $\alpha$ and $\beta$, in order to construct the perturbative Faraday tensor. We begin by recalling that the relationships between $(\gamma_{1},\gamma_{2})$ and $(\alpha, \beta)$ are
 \begin{align}
 & \gamma_1 =\partial_x \alpha +i m \beta\,, \\
 & \gamma_2 = -i m \alpha + (1-x^2) \partial_x [(1-x^2)\beta]\,.
 \end{align} 
If we define $\hat \beta \equiv \beta (1-x^2)$, then these expressions lead to  
 \begin{align}
 & \frac{\partial}{\partial x} \left [ (1-x^2)\frac{\partial \alpha}{\partial x}\right ] -\frac{m^2 \alpha}{1-x^2} = \frac{\partial [(1-x^2)\gamma_1]}{\partial x} -\frac{i m \gamma_2}{1-x^2}\,,\label{eq:AlphaFromGamma}\\
 & \frac{\partial}{\partial x} \left [ (1-x^2)\frac{\partial \hat \beta}{\partial x}\right ] -\frac{m^2 \hat \beta}{1-x^2} = {\frac{\partial \gamma_2}{\partial x}+i m \gamma_1}\,.
\label{eq:BetaFromGamma}
  \end{align}
It is obvious from the form of Eqs.~\eqref{eq:AlphaFromGamma} and \eqref{eq:BetaFromGamma} that $\alpha$ and $\beta$ can be obtained by integrating a Green's function against the sources that appear on the  right hand side, which are linear in $\gamma_{1}$ and $\gamma_{2}$. In other words, $\alpha$ and $\beta$ can be written as 
  \bea
   \alpha = \mathcal{L}^1_{\alpha}(\gamma_1)+\mathcal{L}^2_{\alpha}(\gamma_2), \quad
  \hat{\beta} = \mathcal{L}^1_{\hat{\beta}}(\gamma_1)+\mathcal{L}^2_{\hat{\beta}}(\gamma_2)\,. \label{eq:betafromgamma} 
      \eea
where $\mathcal{L}^{1,2}_{\alpha}$ and $\mathcal{L}^{1,2}_{\hat{\beta}}$ are linear operators acting only on the $x$ coordinate. 

If we specialize to the separable solutions in the form of
\begin{align}\label{eqchiexp}
&\gamma_1 =e^{im\phi}\chi_1(t, r)P_l^m(x),\,\nonumber \\
 &\gamma_2 = e^{i m \phi}(1-x^2)\chi_2(t, r)P_l^m(x)\,,
\end{align}
we can explicitly evaluate the $\mathcal{L}^i_{\cdot}$ operators. First of all, we invoke the following recurrence relation of the associated Legendre polynomials  
\bea
\frac{d[(1-x^2)P^m_l(x)]}{d x} &=& \frac{1}{2l+1} \left [ (l-1)(l+m)P^m_{l-1} \right. \notag \\
&& \left. -(l+2)(l-m+1)P^m_{l+1}\right ]\,
\eea
and the fact that $\gamma_1 = e^{-i\omega t}e^{im\phi}\chi_1(r)P_l^m(x)$ to obtain 
 \bea \label{eq:L1}
 \mathcal{L}^1_{\alpha}(\gamma_1) &=& {e^{-i\omega t} e^{im\phi}}\chi_1 ( r)  \frac{1}{2l+1} \left [ -\frac{(l+m)}{l}P^m_{l-1}\right. \notag \\
 && \left. +\frac{(l-m+1)}{l+1}P^m_{l+1}\right ]\,.
 \eea
 It is also easy to show that 
\begin{equation}  \label{eq:L1beta}
\mathcal{L}^1_{\hat{\beta}}(\gamma_1) =    -{e^{-i\omega t} e^{im\phi}}\chi_1( r) \frac{i m}{l (l+1)} P^m_l = -\frac{i m}{l(l+1)} \gamma_1\,.
\end{equation}
Similarly it is then straightforward to find that 
\begin{equation} \label{eq:L2alpha}
\mathcal{L}^2_{\alpha}(\gamma_2) = \frac{i m \gamma_2}{(1-x^2)\,l\,(l+1)}\,,
\end{equation}  
while $\mathcal{L}^2_{\hat{\beta}}(\gamma_2)$ has the same expression as $\mathcal{L}^1_{\alpha}(\gamma_1) $ but with $\chi_1$ replaced by $\chi_2$. 

\section{Terms in the perturbation equations in the $\alpha$ and $\beta$ basis \label{sec:QandH}}
In order to compute the explicit forms for $\mathcal{Q}_i$ and $\mathcal{H}_i$, we first note that 
the background current associated with the Blandford-Znajek solution is (to $\mathcal{O}(a)$)
\begin{align}
d * F = {-}\Omega_H \cos \theta d \cos \theta \wedge d \phi \wedge du\,, 
 \end{align}
and that the Hodge duals to various two forms under our slow-spinning metric \eqref{eq:slowspinmetric} are
\begin{align} \label{eq:HodgeDuals}
& * dx \wedge d \phi = -\frac{1}{r^2} dr \wedge dt -\frac{1}{r^2} \frac{2a M \sin^2 \theta}{r-2M} dr \wedge d\phi, 
\nonumber \\
 & * dx \wedge d t = -\sin^2\theta \frac{r}{(2M-r)} d\phi \wedge dr + \frac{2 a M \sin^2\theta}{r^2(2M-r)} dt \wedge dr, 
\nonumber \\
 & * dx \wedge dr = -\sin^2\theta \left( 1- \frac{2M}{r} \right) dt \wedge d\phi \nonumber \\
&* dt \wedge d \phi = \frac{r}{(2M-r)\sin\theta} d\theta \wedge dr \,, \nonumber \\
& * dr \wedge d \phi =\left( 1-\frac{2M}{r}\right) \frac{1}{\sin\theta} dt \wedge  d\theta + \frac{2 a M \sin\theta}{r}d\phi \wedge d\theta \,, \nonumber \\
& * dr \wedge d t = - r^2 \sin\theta d\theta \wedge  d\phi + \frac{2 a M \sin\theta}{r}d\theta \wedge dt \,. 
 \end{align}
It is then a tedious but straightforward matter to compute $\delta F$ according to Eq.~\eqref{eq:PertFaraday}, and subsequently the explicit forms of the two equations \eqref{eqwaveab1} and \eqref{eqwaveab2}, which allows us to extract the following expressions:
\begin{widetext}
\bea 
\mathcal{Q}_1 \alpha&=& \sin \theta\left [\Omega_H\cos \theta \left ( \frac{\partial \alpha}{\partial r}+\frac{1}{1-2M/r}\frac{\partial \alpha}{\partial t}\right )+\frac{1}{r^2}\frac{\partial^2 \alpha}{\partial x \partial\phi}-\frac{\partial^2 \alpha}{\partial x \partial t} \frac{(\Omega_Z-\Omega_H/2) \sin^2\theta}{(1-2M/r)}+\frac{\Omega_H \sin^2\theta}{2}\frac{\partial^2 \alpha}{\partial x \partial r} \right ]\,, \label{eq:Q1} \\
\mathcal{H}_2 \alpha  
&=&  \left \{\frac{1}{\sin\theta (1-2M/r)}\frac{\partial ^2 \alpha}{\partial t^2}{-}\frac{\partial^2 \alpha}{\partial x^2}\frac{\sin\theta}{r^2}-\frac{1}{\sin \theta}\left [ (1-2M/r)\frac{\partial \alpha}{\partial r}\right ]_{, r} \right \}  {+ \frac{\Omega_H}{\sin\theta} \left (\frac{1}{1-2M/r} \frac{\partial^2 \alpha}{\partial \phi \partial t}+\frac{\partial^2 \alpha}{\partial \phi \partial r}
\right )  }\,, \label{eq:H2} \\
\mathcal{H}_1 \beta &=&\frac{\sin^3 \theta}{1-2M/r}  \left ( -\frac{\partial^2 \beta}{\partial t^2} +\frac{\partial^2 \beta}{\partial r^2_*}\right ){+}\frac{\sin\theta}{r^2} \frac{\partial^2 \beta}{\partial \phi^2} -2 \Omega_Z \frac{\sin^3 \theta}{1-2M/r} \frac{\partial^2 \beta}{\partial \phi \partial t}
\,, \label{eq:H1}\\
\mathcal{Q}_2 \beta &=&  \frac{1}{r^2} \frac{\partial^2 \beta}{\partial \theta \partial \phi}
+\frac{\Omega_H/2-\Omega_Z}{1-2M/r} \left [ \sin^2 \theta\frac{\partial \beta }{\partial t} \right]_{, \theta} +\frac{\Omega_H}{2} \left [ \sin^2 \theta\frac{\partial \beta }{\partial r} \right]_{, \theta} 
 +\Omega_H\cos \theta \sin \theta \left ( \frac{\partial \beta}{\partial t} \frac{1}{1-2M/r} +\frac{\partial \beta}{\partial r}\right ) \,
  \nonumber \\
 &=&  \frac{1}{r^2} \frac{\partial^2 \beta}{\partial \theta \partial \phi} - \sin(\theta)\left[- \Omega_H\frac{\cos \theta}{\sin^2\theta} \left ( \frac{\partial \hat{\beta}}{\partial t} \frac{1}{1-2M/r} +\frac{\partial \hat{\beta}}{\partial r}\right ) -\frac{\Omega_Z - \Omega_H/2}{1-2M/r}\frac{\partial^2 \hat{\beta}}{\partial t \partial x}
+ \frac{\Omega_H}{2}\frac{\partial^2 \hat{\beta}}{\partial r \partial x}
\right] \,.\label{eq:Q2}
\eea
\end{widetext} 
 
\section{Terms in the perturbation equations in the $\gamma_1$ and $\gamma_2$ basis \label{sec:PerEqGamma}}
In this section, we compute the terms in the perturbation equations \eqref{eqwg1g21} and \eqref{eqwg1g22} that are written under the $\gamma_1$ and $\gamma_2$ basis. We first note that the rescaled perturbation equations
\bea
\left(1-\frac{2M}{r}\right)\frac{\mathcal{H}_1 \beta+\mathcal{Q}_1 \alpha}{\sin^3[\theta]}&=&0,\label{eqwaveeqo1}\\ 
-\left(1-\frac{2M}{r}\right)\sin\theta\left(\mathcal{H}_2 \alpha+\mathcal{Q}_2 \beta \right)&=&0\,, \label{eqwaveeqo2}
\eea
reduces to Eqs.~\eqref{eq:SchCouple1} and \eqref{eq:SchCouple2} in the $a \rightarrow 0$ limit. Then carrying out (on Eqs.~\ref{eqwaveeqo1} and \ref{eqwaveeqo2}) the same basis transformation procedures that lead us from Eqs.~\eqref{eq:SchCouple1} and \eqref{eq:SchCouple2} into Eqs.~\eqref{eq:DiaSch1} and \eqref{eq:DiaSch1}, we obtain Eqs.~\eqref{eqwg1g21} and \eqref{eqwg1g22}, whereby the $\mathcal{H}_V$ and $\mathcal{H}_P$ operators are exactly the same as the operators appearing in Eqs.~\eqref{eq:DiaSch1} and \eqref{eq:DiaSch2}, respectively. The explicit forms of the remaining operators are given by (after substituting $\partial_t$ with $-i\omega$ and $\partial_{\phi}$ with $im$, as we will only need the explicit forms of these operators when applied to separable solutions)
\begin{widetext}
\bea \label{eq:PertTrappedFull}
\mathcal{V}_1 \gamma_1+ \mathcal{V}_2 \gamma_2 &=&
-\frac{1}{8M^2}\left[ -2 i \omega \hat{\beta}
+2\left( 1-\frac{2M}{r}\right)\frac{\partial \hat{\beta}}{\partial r}
+32 m \omega \frac{M^3}{r^3}\gamma_1 
\nonumber \right. \\ && \left. 
- i \omega  \frac{2x}{1-x^2} \gamma_2 
- i \omega \left(16\frac{M^3}{r^3}-1\right)\frac{\partial \gamma_2}{\partial x}
+ \frac{2x}{1-x^2}\left(1-\frac{2M}{r}\right)\frac{\partial \gamma_2}{\partial r}
- \left(1-\frac{2M}{r}\right)\frac{\partial^2 \gamma_2}{\partial r \partial x}
\right]\,,
\eea
and 
\bea
\mathcal{P}_1 \gamma_1+ \mathcal{P}_2 \gamma_2 
&=& 
\frac{1}{8M^2 }\left[
-2 i \omega (1-x^2)\alpha
+2\left(1-\frac{2M}{r} \right)(1-x^2)\frac{\partial \alpha}{\partial r}
\nonumber \right. \\ && \left. 
-32 i \omega \frac{M^3}{r^3} x \left(1-x^2\right) \gamma_1  
+i \omega (1-x^2)^2\left(16\frac{M^3}{r^3}-1\right)\frac{\partial \gamma_1}{\partial x}
+\left(1-\frac{2M}{r}\right)(1-x^2)^2 \frac{\partial^2\gamma_1}{\partial r \partial x}
\nonumber \right. \\ && \left. 
-2 m \omega \gamma_2 - 2 i m \left(1-\frac{2M}{r}\right)\frac{\partial\gamma_2}{\partial r}
\right]\,,
\eea
\end{widetext}
where $\hat\beta$ and $\alpha$ should be seen as implicit functions of $\gamma_1$ and $\gamma_2$ through the $\mathcal{L}_{\cdot}^i$ operators in Eq.~\eqref{eq:betafromgamma}. 
When $\gamma_1$ is restricted to its separable Schwarzschild limits $\gamma^{(0)}_1$, such as when we compute $\delta \omega$ according to Eq.~\eqref{eq:FreqShift}, we can replace the $\mathcal{L}_{\hat\beta}^1$ operator with its explicit expression 
from Eq.~\eqref{eq:L1beta}, and obtain (when acting on $\gamma^{(0)}_1$)
\bea \label{eq:Integrands}
\mathcal{V}_1 &=& \frac{m}{8M^2}\left[ 
\left( 1-\frac{2M}{r}\right)\frac{2i}{l(l+1)}\frac{\partial}{\partial r} \right. \notag \\
&& \left. - 32  \omega_S \frac{M^3}{r^3} +\frac{2\omega_S}{l(l+1)}
\right]\,.
\eea
Similarly, when $\gamma_2$ is assumed to have the separable form of Eq.~\eqref{eqg2c2} (not necessarily restricted to $\gamma^{(0)}_2$), we can use Eq.~\eqref{eq:L2alpha} to obtain 
\bea \label{eq:P2}
\mathcal{P}_2 = \frac{m}{4M^2}\left(\frac{1}{l(l+1)}-1\right)\left(\omega + i\frac{\partial}{\partial r_*} \right)
\eea

\bibliography{References} 


\end{document}